\documentclass[preprint,aps,pre,groupedaddress,showpacs,preprintnumbers,amsmath,amssymb]{revtex4}
\usepackage{bm,amsmath,amsthm}
\usepackage{graphicx}
\usepackage{subfigure}

\newcommand\eq[1]{\begin{eqnarray}#1\end{eqnarray}}

\newcommand{\inte}{\!\int\!\!}
\def\dbar{{\mathchar'26\mkern-12mu d}}
\newcommand{\ud}{\dbar}

\newcommand{\mar}{\bm r}
\newcommand{\maxx}{\bm x}

\newcommand{\maq}{\bm q}
\newcommand{\map}{\bm p}

\begin{document}
\title{Steady state existence of passive vector fields under the Kraichnan model}
\author{Heikki Arponen}
\affiliation{Helsinki University, Department of Mathematics and
Statistics, P.O. Box 68, 00014 Helsinki (Finland)}
\email{heikki.arponen@helsinki.fi}
%
%
%
%
%
%
\date{\today}
\begin{abstract}
The steady state existence problem for Kraichnan advected passive
vector models is considered for isotropic and anisotropic initial
values in arbitrary dimension. The models include the
magnetohydrodynamic (MHD) equations, linear pressure model (LPM)
and linearized Navier-Stokes (LNS) equations. In addition to
reproducing the previously known results for the MHD model, we
obtain the values of the Kraichnan model
roughness parameter $\xi$ for which the LNS steady state exists.
\end{abstract}
\pacs{47.27.E-, 47.27.-i}
\maketitle
%
%
%
%
\section{Introduction}
This is a companion paper to a previous work by the present author
\cite{arponen2}, wherein the phenomenon of anisotropic anomalous
scaling was studied in the context of passive vector fields. The
work was in part incomplete, as the main assumption was the
existence of a steady state solution for the pair correlation
function. We aim here to find exactly the preconditions under
which this assumption is valid. Much of the technical material is
from the above paper, to which we often refer for details. We
study the stability of an equal time pair correlation function of
a field $\mathbf{ u} (t,x)$ determined by the equation
\eq{ \dot u_i - \nu \Delta u_i + \mathbf v  \cdot \nabla u_i - a
\mathbf u \cdot \nabla v_i  + \nabla_i P = 0 ,\label{equation}}
where the vector field $\mathbf{v} (t,x)$ is determined by the
Kraichnan model \cite{kraichnan} and all vector quantities are
divergence free.

The model was introduced in \cite{paolo} as the most general
linear passive vector model respecting galilean invariance. The
parameter $a = -1,0$ or $1$ corresponds respectively to the
linearized Navier-Stokes equations \cite{paolo} (abbreviated
henceforth as LNS), the so called linear pressure
model\cite{paolo,arad3,adz,benzi,jurc1,nikolai} (LPM) and the
magnetohydrodynamic (MHD) equations
\cite{paolo,arad,antonovlanottemazzino,vergassola,arponen,jurc2,lanotte2,vincenzi,kazantsev,nikolai2}.
In the context of the magnetohydrodynamic case, the inexistence of
the steady state is known as the "dynamo effect", where the dynamo
refers to exponential growth in time of the pair correlation
function (see e.g. \cite{arponen,vincenzi,kazantsev} and
references therein). This problem is by far the easiest of the
three due to the vanishing of the nonlocal pressure effects. In
\cite{arad3} it was shown that for the linearized pressure model
(corresponding to $a=0$) the steady state always exists by showing
that the semi-group involved with the time evolution is always
positive. The analysis for the linearized Navier-Stokes case with
$a=-1$ is considerably more difficult than the above two cases
since unlike in the MHD case, the nonlocal effects are present and
contribute strongly to the dynamics, and because unlike in the LPM
case, the semigroup is not always positive.

The present goal is therefore to find the values of $\xi$ for
which the LNS steady state exists, where $\xi$ is the roughness
exponent of the Kraichnan velocity field $v \sim r^\xi$. The
method by which this is accomplished involves applying a Mellin
transform on the eigenvalue equation for the two point correlation function,
and by solving a resulting recursion relation.
\section{The model}
We sketch here the derivation of the equation in Mellin
transformed form and refer to the previous paper \cite{arponen2}
for further details. All vector quantities in eq. (\ref{equation})
being divergence free results in an expression for the pressure,
\eq{P = (1-a) \left( - \Delta\right)^{-1} \partial_i v_j
\partial_j u_i.}
One may then rewrite the equation compactly as 
\eq{\dot u_i - \nu \Delta u_i  + \mathcal D_{ijk} \left( u_j v_k
\right) = 0,}
with an integro-differential operator
\eq{\mathcal D_{ijk} = \delta_{ij} \partial_k - a \delta_{ik}
\partial_j + (a-1)\partial_i \partial_j \partial_k \Delta^{-1},}
where $\Delta^{-1}$ is the inverse laplacian. The equal time pair
correlation is defined as
\eq{G_{ij}(t,\mar) = \langle u_i (t,\maxx+\mar) u_j (t,\maxx)
\rangle,}
where the angular brackets denote an ensemble average with respect
to the velocity field, which in turn is defined by the Kraichnan
model as
\eq{ \big\langle  v_i(t,\mar)v_j(0,0)\big\rangle &&= \delta(t)
D_{ij}(\mar) \nonumber \\ &&= \delta(t) D_1  \inte \ud^d \maq
\frac{ e^{i \maq \cdot
\mar}}{\left(\maq^2+m_v^2\right)^{d/2+\xi/2}} P_{ij}(\maq)
\label{KraichCorr}}
where we have defined the incompressibility tensor
$P_{ij}(\maq)=\delta_{ij}-\widehat \maq_i \widehat \maq_j$, $\ud^d \maq := \frac{d^d q}{(2 \pi)^d}$ and $\xi \in [0,2]$ is a parameter describing the spatial roughness of the flow.

We note a subtle difference from \cite{arponen} in that we define the constant
\eq{D_1 = \frac{4 \xi \Gamma \left(  \frac{2+d+\xi}{2}
\right)}{\Gamma \left( 1-\xi/2 \right)} D_0.}
The reason for this is that the velocity correlation and structure
functions would otherwise diverge at $\xi=0$ and $\xi=2$ as the
mass cutoff is removed. This aspect of the Kraichnan model has been clearly discussed in \cite{paolo.antti}. The equation for the pair correlation
function is then
\eq{\partial_t G_{ij} -2 \nu \Delta G_{ij} - \mathcal D_{i\mu \nu}
\mathcal D_{i\rho \sigma} \left( D_{\nu \sigma} G_{\mu \rho}
\right) = 0.\label{paircorr}}
The Fourier transform of the correlation function will then be
decomposed in terms of hyperspherical tensor basis according to
the prescription in \cite{arad2} as
\eq{\widehat{G}_{ij} (t, \map) := \sum\limits_{a,l} B_{ij}^{a,l}
(\hat \map) \widehat G_{a,l} (t, p) \label{covariances},}
where the tensor basis components are
\eq{  \ \left\{ \begin{array}{ll}
 B_{ij}^{1,l} (\hat \map) &= |\map|^{-l} \delta_{ij} \Phi^l (\map) \\
 B_{ij}^{2,l} (\hat \map) &= |\map|^{2-l} \partial_i \partial_j \Phi^l
(\map)
 \\
 B_{ij}^{3,l} (\hat \map) &= |\map|^{-l} (p_i \partial_j + p_j
\partial_i)
 \Phi^l (\map) \\
 B_{ij}^{4,l} (\hat \map) &= |\map|^{-l-2} p_i p_j \Phi^l (\map)
\end{array} \right. }
and where $\Phi^l (\map)$ is defined as $\Phi^l (\map) := |\map|^l
Y^l (\hat \map)$, where $Y^l$ is the hyperspherical harmonic
function (with the multi-index $m=0$). It satisfies the properties
\eq{\Delta \Phi^l (\map) &&= 0 \nonumber \\ \map \cdot \nabla
\Phi^l (\map) &&= l \Phi^l (\map).\label{phiproperties}}
Note that we are concerned only with even parity and axial
anisotropy. We now introduce the Mellin transform which will be used to transform
the equation into a recurrence/differential equation. The method was (probably) first used in \cite{Barnes} in the context of the hypergeometric function. The textbook by Hille \cite{Hille} also has a useful section on
the Mellin transform applied to differential equations. Appendix \ref{appendix1} of the present work also contains helpful material, and also offers some insight into some limitations of the method. We define the Mellin transform of a Fourier transform of $G$
(the anisotropy index $l$ will usually be omitted) as
\eq{\bar g_a (t, z) = \int\limits_0^\infty \frac{dw}{w} w^{d+z}
\widehat G_{a} (t, w)\label{mellin}}
and the inversion formula
\eq{G_a(t, \mar) = \int_{0-} \ud z |\mar|^{z} \mathcal
A_z \bar g_a (t, z)\label{f=Ag}}
with the definition $\mathcal A_z =
\frac{\Omega_d}{(2 \pi)^d} \frac{\Gamma \left( d/2 \right) \Gamma
\left( -z/2 \right)}{2^{z+1} \Gamma \left( \frac{d+z}{2} \right)}$, which
originates from the inversion of the fourier integral (with volume
of the unit sphere $\Omega_d$), and the subscript $0-$ was used to denote a contour from $- \imath \infty$ to $+ \imath \infty$ passing $z=0$ from the left (see appendix \ref{appendix1} for details). We also often denote $\bar f (z) \doteq \mathcal A_z \bar g (z)$, which is just the ordinary Mellin transform.

Applying the Fourier transform, dividing by $p^2$, applying the
Mellin transform and finally setting the cutoff parameter $m_v$ to
zero in eq. (\ref{equation}) (see \cite{arponen2} for details), we
obtain the complex recurrence/differential equation
\eq{\partial_t \bar g_a (t,z-2) + 2 \nu \bar g_a (t,z) -
\widetilde \lambda \bar g_a (t,z-\xi) - \mathrm
T^{ab}_{d+\xi,d+z-\xi} \bar g_b (t,z-\xi)= 0,}
with the definitions
\eq{&&\widetilde \lambda = (a-1)\left(d+1+a(1-\xi)\right)
 \frac{d \pi \xi \Gamma
(d/2) c_d}{16 \sin(\pi \xi/2) \Gamma \left( \frac{d-\xi}{2}+2 \right) \Gamma
\left( \frac{d+\xi}{2}+1 \right)}\label{lambda}}
and
\eq{\mathrm T^{ab}_{2\alpha,2\beta} = \frac{4 \xi \Gamma \left(
\frac{d+\xi}{2} \right)}{\Gamma \left( 1-\xi/2 \right)}
\frac{\Gamma (d/2+l-\alpha)
 \Gamma (d/2-\beta) \Gamma (\alpha+\beta-d/2)}{\Gamma
  (\alpha) \Gamma (\beta) \Gamma (d+l-\alpha-\beta)} \tau^{ab} (z),}
where the matrix coefficients $\tau^{ab} (z)$ are listed in the
appendix of \cite{arponen2}. We have also effectively set $D_0 =
1$ by redefining time and viscosity. Requiring the correlation
function (\ref{covariances}) to be divergence free, i.e. zero when
contracted with $p_i$, results in only two of the four
coefficients $\bar g_a$ being independent. The resulting equation
may then be written in the following form,
\eq{\partial_t \mathbf{\bar h} (t,z+\xi-2) + 2 \nu \mathbf{\bar h}
(t,z+\xi) -\left(\widetilde \lambda \mathbf{1} + \mathbf A+
\mathbf B \cdot \mathbf X \right) \mathbf{\bar h} (t,z) =
0.\label{eq2}}
Here we have performed a translation $z \to z +\xi$, defined the
vector quantity $\mathbf{\bar h} = \left(\bar g_1 ,\bar g_2
\right)^{\mathrm T}$
and the matrices by
\eq{ \mathbf T_{d+\xi,d+z} =
\begin{pmatrix}
 \mathbf A & \mathbf B \\ \mathbf C & \mathbf D
\end{pmatrix} \ \ \ , \ \ \ \mathbf X = \begin{pmatrix}
 0 & -(l-1) \\
 -1 & l(l-1)
\end{pmatrix}.}
\subsection{Isotropic sector}
We will be mostly concerned with the isotropic case since much of
the actual computations can be neatly performed all the way. For
$l=0$, only the tensors $B^1$ and $B^4$ are nonzero, and
correspondingly in the tensor decomposition we only have the
coefficients $\bar g_1 (z)$ and $\bar g_4 (z)= - \bar g_1 (z)$
(due to the divergence free condition). The equation (\ref{eq2})
then becomes a scalar equation for $\bar g_1 (z)$ alone, hence we
only need the $(1,1)$ component of the matrix $\left(\widetilde
\lambda \mathbf{1} + \mathbf A+ \mathbf B \cdot \mathbf X
\right)$, which reads explicitly
%
\begin{widetext}
\eq{&&\left(\widetilde \lambda \mathbf{1} + \mathbf A+ \mathbf B
\cdot \mathbf X \right)_{11} =  \frac{2(a-1) (a \xi -1 - a - d)
\Gamma \left( 1+\xi/2 \right) \Gamma \left( 1+d/2 \right)}{\Gamma
\left( \frac{4+d-\xi}{2} \right)}
\nonumber \\
&& -p_a(z) \frac{\Gamma \left( -z/2 \right)  \Gamma \left(
\frac{d+z+\xi}{2} \right)
  }{2 \Gamma \left(
  \frac{2+d+z}{2} \right)
   \Gamma \left( \frac{4-z-\xi}{2} \right) }
   \doteq \Lambda^a_\xi (z),\label{1overgamma}}
\end{widetext}
where
\eq{&&p_a(z) = -(a-1)^2 (d+1)\xi (2-\xi) \nonumber \\ &&
+(z+\xi-2) \left((d-1) z^2 +\left( d(d-1)+2 a \xi \right)z +
\right. \nonumber \\ &&\left. \xi \left( -d-1+2 a
(d+1)-a^2(1+2d-d^2+\xi-d \xi) \right) \right)}
The equation in the isotropic sector is then
\eq{\partial_t \bar g_1 (t,z+\xi-2) + 2 \nu \bar g_1 (t,z+\xi)
-\Lambda^a_\xi (z) \bar g_1 (t,z) = 0.\label{isotropiceq}}
\section{The method}
%
%
%
%
We now consider the eigenvalue problem with $\bar g_1 (t,z)
\propto e^{-E t} g (z)$, resulting in the equation
\eq{E \bar g (z+\xi-2) - 2 \nu \bar g (z+\xi) +\Lambda^a_\xi (z)
\bar g (z) = 0.\label{eigen}}
This is analogous to the Schr\"odinger method in
\cite{vergassola,vincenzi,kazantsev}. The steady state exists if
one can show that the spectrum is nonnegative. However, for
example in the magnetohydrodynamic case as in the above mentioned
papers and in \cite{arponen}, it was shown that there exists a
critical value of the parameter $\xi$ above which one has negative
energies resulting in an exponential growth in time. This
phenomenon is interpreted as the dynamo effect of magnetic fields.
Previous studies on the dynamo problem have resorted to some
approximative or numerical schemes to find the growth rate $|E|$
as a function of $\xi$. Here we will settle for simply finding the
values of $\xi$ for which the energies are nonnegative, thus
implying a steady state. This is done by studying the
\textit{zero energy} equation, i.e. setting $E=0$. This has the
advantage of providing us with an exact solution, up to a
numerical solution of a transcendental equation.

The argument used in the present work is closely analogous to the classic "node theorem" (see e.g. \cite{reedsimon} and also appendix \ref{appendix1}), which can be roughly stated as follows. Suppose that for some large enough value of $E$ the corresponding solution $f_E (r)$ is oscillating between positive and negative values ar large $r$ and satisfies the boundary condition $f_E (L)=0$, for some large $L$ (tending to infinity). If the spectrum is bounded from below, we know that by decreasing $E$ the zeros of $f_L (r)$ will move to the right and satisfy the boundary condition for a discrete set of $E$. The value of $E$ for which the smallest zero reaches $L$ is then the ground state energy. Since we are interested in whether or not the ground state is positive or negative, we can instead study the zero energy equation and ask for which values of $\xi$ the solution crosses over from nonoscillating to oscillating. We know that the $\xi=0$ equation corresponds to diffusion, i.e. a nonoscillating zero energy ground state. As we increase $\xi$, we may discover that the solution becomes oscillating at large scales, which would imply a negative energy ground state and therefore instability. The large scale behavior of the solutions is determined by the negative poles of the Mellin transform, so the problem is then reduced to finding these poles and determining if and when they become complex valued.
\section{Isotropic sector of LNS}
\begin{figure*}[h]
\begin{center}
\includegraphics[scale=.65]{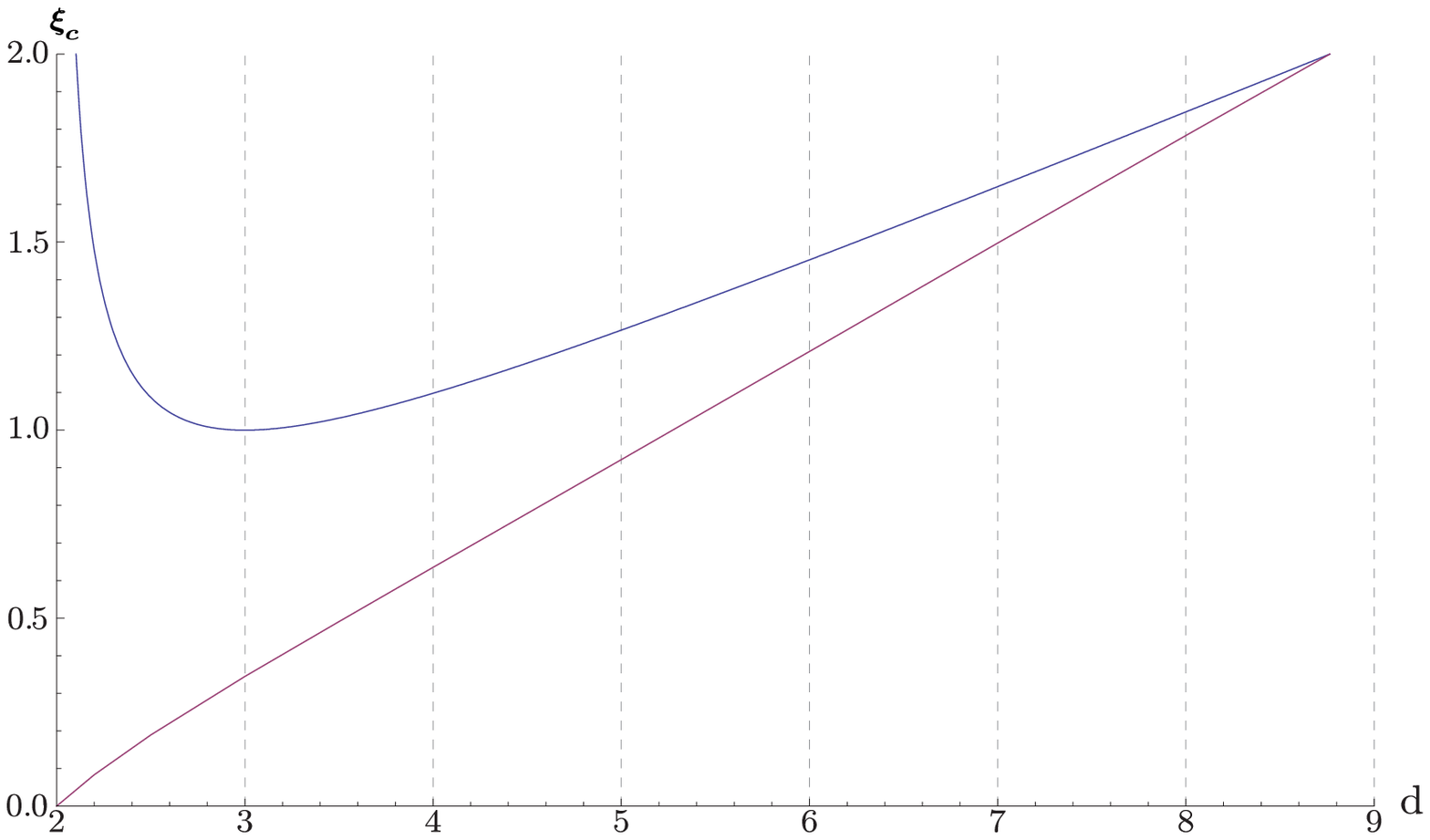}
\caption{ A plot of the critical value $\xi_c$, above which the
flow becomes unstable, versus the space dimension $d$. The upper
curve refers to the magnetohydrodynamic case with $a=1$ and the
lower to the linearized Navier-Stokes case with
$a=-1$.\label{criticalxi}}
\end{center}
\end{figure*}

The stability problem in the linearized Navier-Stokes case is
closely related to the laminar flow stability problem as described
in \S 26 of \cite{landau}. The equation is derived from the
Navier-Stokes equation by decomposing the velocity field into
$\mathbf{v}(\mar) + \mathbf{u} (t,\mar)$, where $\mathbf{v}$ is a
stationary solution and $\mathbf{u}$ is a small perturbation,
resulting in the equation
\eq{ \dot u_i - \nu \Delta u_i + \mathbf v  \cdot \nabla u_i +
\mathbf u \cdot \nabla v_i  + \nabla_i P = 0 .}
The question is then whether or not the laminar flow is stable
under such perturbations. Here instead the field $\mathbf{v}$ is
supposed to model a \textit{statistical steady state} solution of
the full Navier-Stokes turbulence, as prescribed by the Kraichnan
model. We are therefore studying whether or not the Kraichnan
model is an adequate steady state description of turbulence in
terms of stability. For $a=-1$ we now have
\begin{widetext}
\eq{&&\Lambda^{-}_\xi (z) =  \frac{4 (d+\xi) \Gamma \left( 1+\xi/2
\right) \Gamma \left( 1+d/2 \right)}{\Gamma \left(
\frac{4+d-\xi}{2} \right)} -p_{-1}(z) \frac{\Gamma \left( -z/2
\right)  \Gamma \left( \frac{z+d+\xi}{2} \right)
  }{2 \Gamma \left(
  \frac{z+d+2}{2} \right)
   \Gamma \left( \frac{4-\xi-z}{2} \right) }
   ,\label{isotropic.poles}}
\end{widetext}
where
\eq{p_{-1} (z) &&=(z+\xi)  \left( -2 z+2 \xi +d^2 (z+\xi-2 )
\right. \nonumber \\ &&\left. -(z+\xi )^2+d (2+(z-3) z-(3-\xi )
\xi ) \right) +4 z^2.}
The problem is obviously a more difficult one than in the MHD case
due to the transcendental nature of the function $\Lambda_\xi^-$.
We can however expand it as an infinite product of zeros
and poles according to the Weierstrass factorization theorem (see
e.g. \cite{conway}). The function
$\Lambda_\xi^-$ may then be rewritten as
\eq{\Lambda_\xi^- (z) = e^{s(z)}\prod\limits_{k=1}^{\infty}
\frac{\left(
 z-a_k^+ \right) \left( z-a_k^- \right)}{\left( z - b_k^+ \right)
\left( z-b_k^- \right)} e^{\delta /k},}
where the $+$ and $-$ signs refer to zeros or poles that have
respectively positive (or zero) or negative real parts, see Fig.
(\ref{criticalxi}). We also have the poles $b_k^+ = 2 k$, $b_k^- =
-d-\xi -2k$ and $s(z)$ is some unknown entire function on the
complex plane and $\delta$ is a $z$ -independent Weierstrass
factor that enforces convergence of the infinite product. It can
be derived by showing that asymptotically as $z \to \pm \infty$,
the poles and zeros of eq. (\ref{isotropic.poles}) behave
respectively as $\pm 2 k + \text{const.}$ where the constant term
depends on $\xi$ and $d$. It may certainly be possible to derive
bounds for $s(z)$ by asymptotic analysis of eq.
(\ref{isotropic.poles}), but since it can not contribute to the
pole or zero structure of the solution, we refrain from doing so.
We can also neglect the explicit form of the constant $\delta$ for
the same reason. The zero energy equation from (\ref{eigen}),
rewritten here as
\eq{\bar g (z) &&= \frac{2 \nu}{\Lambda_\xi^- (z)} \bar g (z+\xi)
,\label{zeroLNS}}
can then be solved by the same methods as in appendix \ref{appendix1} with the
strip of analyticity requirement $-\xi < \mathcal R e (z) <0$,
resulting in
%
%
%
%
%
%
%
%
%
%
%
\eq{\bar g (z) &&= \sigma_\xi (z) \Psi(z) (2\nu)^{-z/\xi}  \nonumber \\
 &&\times \prod\limits_{k>0} e^{z \delta/\xi k}
 \frac{\Gamma\left( \frac{z-a_k^-}{\xi} \right)
 \Gamma\left( \frac{2k + \xi-2-z}{\xi} \right)}{\Gamma\left( \frac{a_k^+ + \xi - z}{\xi} \right)
 \Gamma\left( \frac{2k+z+d+\xi-2}{\xi} \right)}\label{LNSsolution}}
%
%
%
where $\Psi(z)$ satisfies the equation $\Psi(z) = e^{ -s(z)} \Psi
(z+\xi)$, whose solution is again an exponential of an entire
function. The following subtlety concerning the above formula
should be observed: we deliberately chose to use the form $\sim
1/\Gamma(1-x)$ instead of $\Gamma (x)$, where the two are related by the Euler reflection formula $\Gamma (-z/2) \Gamma(1+z/2) = -\pi / \sin (\pi
z/2)$. The reason is that only in
this form the strip of analyticity remains pole free, as per the consistency requirement. For example
using $\Gamma \left( \frac{z-a_k^+}{\xi} \right)$ in the above
result would introduce poles at $z=a_k^+ - \xi n$ for positive
integers $n$, that would eventually permeate the strip of
analyticity. However, in some cases as we increase $\xi$, the poles will enter the strip of analyticity and render the solution incompatible with the strip of analyticity requirement. We will also demand that the solution converges to zero at imaginary infinities in order to justify the shift of integration contour (see appendix \ref{appendix1}). It seems quite difficult to deduce the asymptotic behavior from the above formula, but we can study it by an asymptotic expansion of the
exact form of eq. (\ref{isotropic.poles}). The function
$\Lambda_\xi^- (z)$ behaves asymptotically as $\sim z^\xi$ at
imaginary infinities, so the asymptotic version of the difference
equation (\ref{zeroLNS}) reads
%
%
\eq{\bar g (z) = z^{-\xi} \bar g (z+\xi)}
up to some irrelevant constant term. The asymptotic solution is
then $\bar g (z) = \sigma_\xi (z) \Gamma (z)^\xi$. Multiplication
by $\Gamma (-z/2) / \Gamma \left( (z+d)/2 \right)$ in defining
$\bar f$ introduces a pole at $z=0$, which takes care of the
boundary condition. Then we have asymptotically $\bar f (z) \sim
\sigma_\xi (z) e^{-\frac{\pi}{\xi} y} y^{(\xi-1)(x-1/2)}$, where
$z=x+\imath y$. Fourier expansion of $\sigma_\xi$ would then
contain terms such as $\sin \left( \frac{2\pi}{\xi} n z \right)$,
which would spoil integrability for $0 < \xi < 2$, unless $n=0$.
Therefore $\sigma_\xi (z)$ has to be a constant.
\begin{figure*}[h]
\begin{center}
\includegraphics[scale=1]{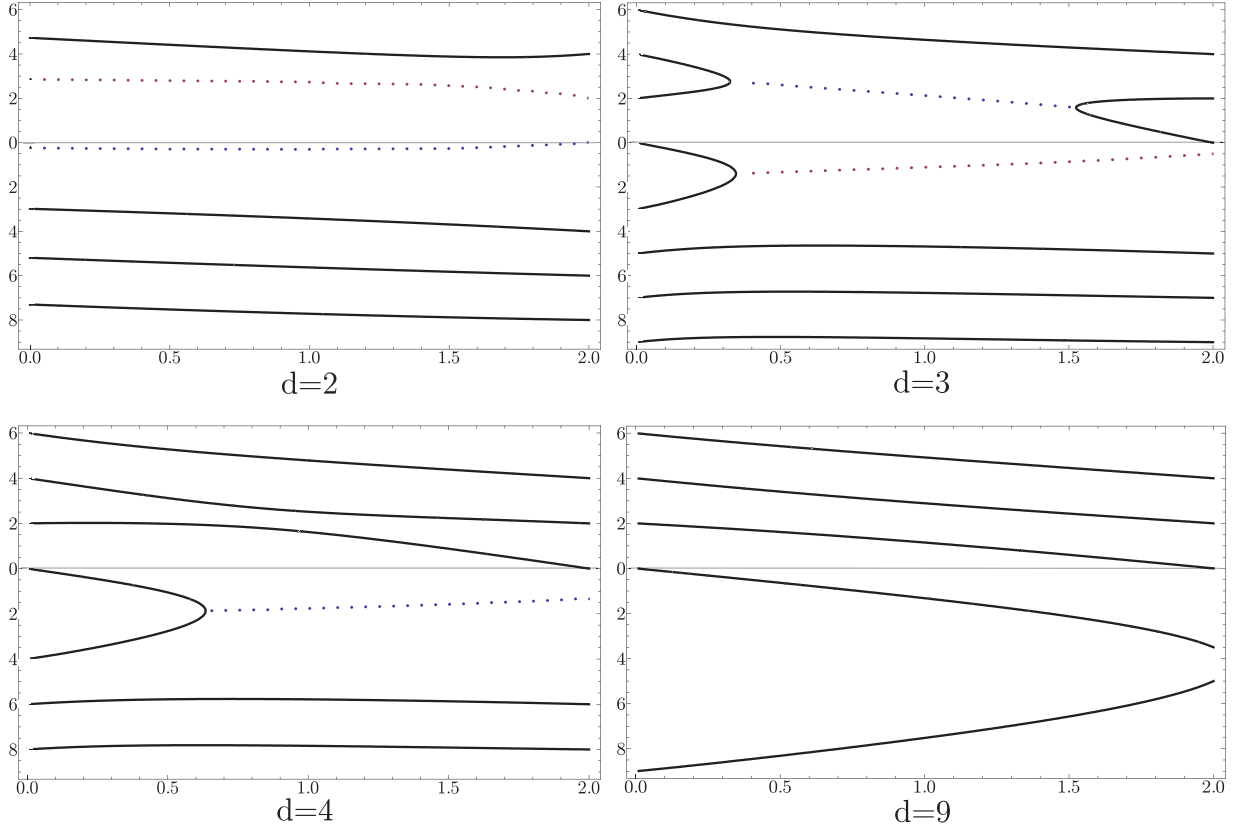}
\caption{A plot of the $a=-1$ poles of the solution $\bar g (z)$
in the isotropic sector versus $\xi$ in various dimensions. The
dashed curves denote the real parts of complex valued poles. There
is an infinity of poles, but all the others not displayed here are
real. Also, all the poles in dimensions $d \ge 9$ are real in $0
\le \xi \le 2$.} \label{poles}
\end{center}
\end{figure*}
We see now that the poles of $\bar f (z)$ occur for non-negative
integers $n$ at $z=\xi -2+2k+\xi n$, $z=2(k-1)$ (with $k>0$) and
at $z= a_k^- - \xi n$, where only the latter affects the large
scale behavior. We draw the important conclusion that
the poles $a_k^+$ have no effect on the steady state
existence problem. By looking at Fig. (\ref{poles}) we can see how
the first few large scale poles $a_k^-$ behave in various
dimensions. In two dimensions the leading pole $a_1^-$ enters the strip at around $\xi \approx 0.28$ and is complex
for all $\xi$, which implies that there is no steady state at all
in the isotropic sector, at least for $\xi < 0.28$ \footnote{The previous paper
\cite{arponen2} by the present author failed to address this
complex valued pole. This led to an incorrect hypothesis about the
existence of the steady state}. In three dimensions the poles
$a_1^-$ and $a_2^-$ become complex at around $\xi_c \approx
0.345$ and enter the strip of analyticity at around $\xi \approx 1.08$. Similar behavior occurs with different $\xi_c$ in higher
dimensions, until at $d=9$ the poles stay real for all $0 \le \xi
\le 2$. We have plotted the value of $\xi_c$ in Fig.
(\ref{criticalxi}) in dimensions $2 \ldots 9$ together with the
magnetohydrodynamic case. The fact that in some cases for large enough $\xi$ the strip of analyticity condition is violated could possibly mean that the steady state exists also for some large values of $\xi$. We will however be content with studying the cases for which such a violation does not take place.

One important lesson of the present section is that the
"complexification" hypothesis of \cite{arponen2,arponen} is indeed
an indication of instability of the flow, in the sense that the
imaginary parts of the scaling exponents correspond to
oscillations of the correlation function and are therefore
responsible for the instability. The second lesson is that one
only needs to be concerned with the negative zeros $a_k^-$
of $\Lambda_\xi^a (z)$ when considering the stability problem,
since they become the poles in the solution $\bar f (z)$. The
positive zeros $a_k^+$ appear only as zeros in $\bar f (z)$.
\section{Anisotropic sectors of LNS}
\begin{figure*}[h]
\begin{center}
\includegraphics[scale=1]{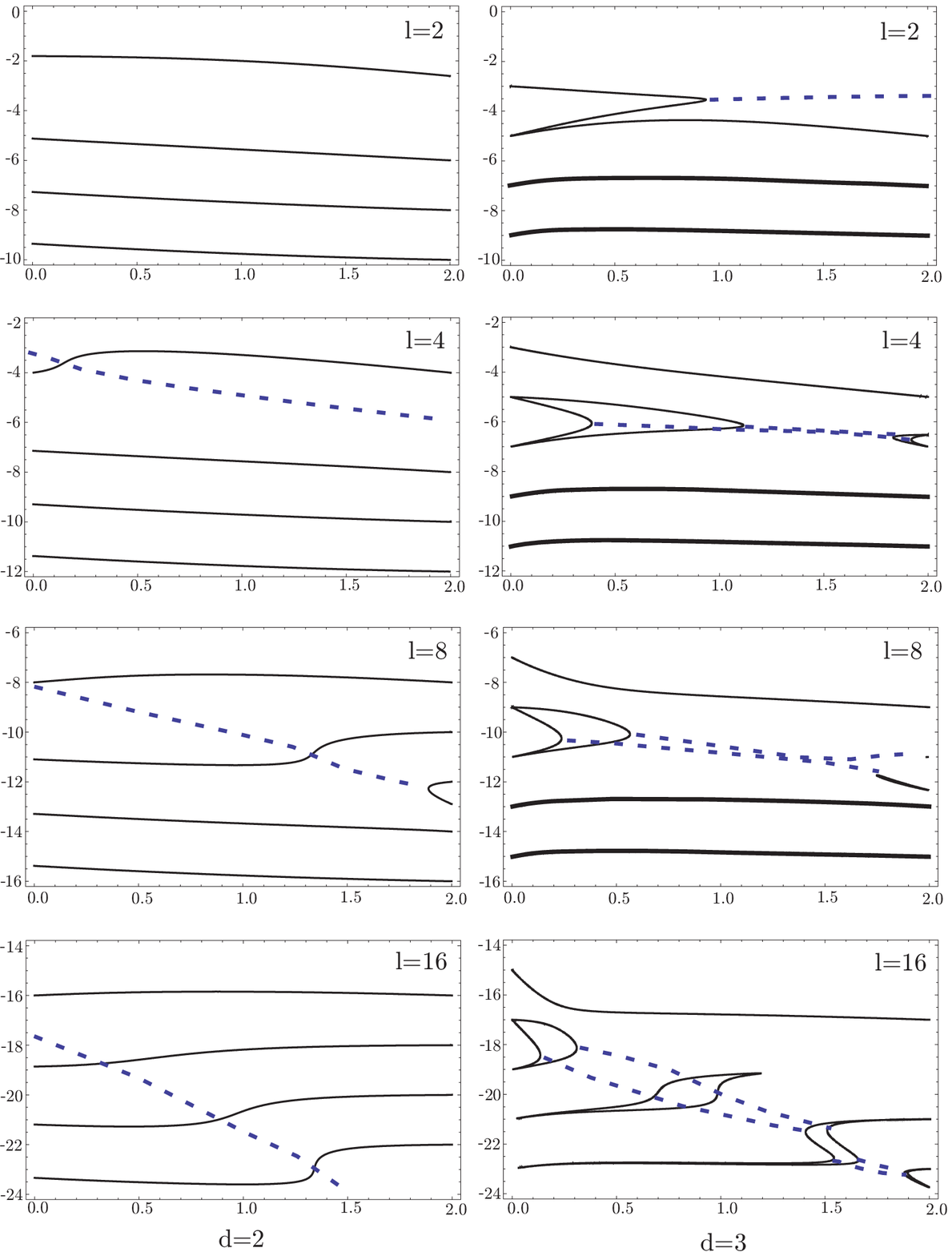}
\caption{A figure showing the leading large scale poles $a_k^-$ in
various anisotropic sectors in two and three dimensions. The
dashed lines denote the real parts of complex valued poles. The
thick curves denote two poles very close to each other. All the
poles beyond what are shown here are real for $0<\xi<2$.}
\label{anisotropic.poles}
\end{center}
\end{figure*}
The anisotropic sectors can be studied with the same methods as
above. We will however refrain from performing the actual
computations and simply extend what we have learned from the
isotropic sector to the anisotropic case, namely that one simply
needs to study the complexification of the negative poles of the
solution $\bar f_a (z)$ (we now have a matrix equation). The role
of $\Lambda_\xi$ will now be taken by the determinant of the
matrix in eq. (\ref{eq2}), instead of just the $(1,1)$ component.
Since we already know that the flow is stable in the anisotropic
sectors for $a=1$ and $a=0$ (see e.g. \cite{arad2,arad3}), we
concentrate only on the $a=-1$ case in various dimensions. The
anisotropic linearized Navier-Stokes exponents differ from the
magnetohydrodynamic ones in that even if the leading exponent is
real, the next to leading exponent may be complex valued,
resulting in oscillating behavior at intermediate scales. These
exponents however have no effect on the existence problem. This is
because the boundary condition at $R$ tending to infinity can only
be satisfied by the leading oscillating exponent. We also need to make sure that the periodic function $\sigma_\xi (z)$ is again required to be a constant due to integrability, so that it will not cancel with any of the poles. This results from the fact that all matrix coefficients beside the (1,1) coefficient in eq. (\ref{1overgamma}) behave asymptotically as $\sim z^\xi$
irrespectively of $l$ \footnote{This can easily be verified by
studying the asymptotics of the matrix coefficients $T^{ab}$ in
appendix C of \cite{arponen2}.}, and therefore so does the
determinant. Hence the same conclusions will be drawn as in the
isotropic case, i.e. that $\sigma_\xi$ is indeed a constant.

We have plotted some leading negative exponents in various
anisotropic sectors in two and three dimensions in fig.
(\ref{anisotropic.poles}). In two dimensions the leading
anisotropic exponent is real, except (strangely enough) for $\xi
\lesssim 0.15$. The anisotropic sectors are therefore quite stable
in comparison to the completely unstable isotropic sector. In
three dimensions the $l=2$ anisotropic leading exponent becomes
complex at $\xi_c^{(2)} \approx 0.937$ and all the higher sectors
have purely real leading exponents. The anisotropic sectors are
therefore much more stable in comparison to the isotropic sector
critical value $\xi_c^{(0)} \approx 0.345$, somewhat similarly to
the magnetohydrodynamic case. We also note that none of the poles
lie inside the strip of analyticity, so the results should hold for all $\xi$. In dimensions $d
\ge 4$ the anisotropic exponents are always real.
\section{Conclusion}
The stability analysis of the passive vector models previously
considered in a companion paper \cite{arponen2} was successfully
completed. The critical value $\xi_c$ below which the steady state
exists was found in all dimensions, although the possibility of a
steady state for an even larger region could not be excluded in
the isotropic sector. The reason for flow instability was shown to
be caused by the complexification of the largest negative pole of
the solution, corresponding to large scale behavior of the
correlation function. It was observed that in two dimensions the
linearized Navier-Stokes problem is not stable for any $\xi>0$ in
the isotropic sector, but relatively stable in the anisotropic
sectors. In three dimensions the isotropic sector was observed to
be stable for $\xi \lesssim 0.345$, the $l=2$ anisotropic sector
for $\xi \lesssim 0.937$ and higher anisotropic sectors for all
$\xi$. In dimensions from four to eight, the isotropic sector is
stable below the critical values plotted in Fig.
(\ref{criticalxi}) and the anisotropic sectors are stable for all
$\xi$. In dimensions $d \ge 9$, all sectors are stable for all
$\xi$.
\begin{acknowledgements}
The author wishes to thank P. Muratore-Ginanneschi and A.
Kupiainen for useful discussions, suggestions and help on the
matter. Especially the computer algebra packages of P. M-G. have been of invaluable help \cite{paolo.packages}. This work was supported by the Academy of Finland
"\emph{Centre of excellence in Analysis and Dynamics Research}"
and TEKES project n. 40289/05 "\emph{From Discrete to Continuous
models for Multiphase Flows}".
\end{acknowledgements}
\newpage
\appendix
\section{Mellin transform and examples \label{appendix1}}
The Mellin transform has been used to solve differential equations previously in e.g. \cite{Hille,Barnes}. The purpose of the present appendix is to clarify some aspects of the Mellin transform method and to point out some of its limitations.
\subsection{Solving differential equations by Mellin transform}
Define the Mellin transform of a function $f(r)$ with $r \geq 0$ as \footnote{Note that this definition differs from eq. (\ref{mellin}), which was defined for the Fourier transform.}
\eq{\bar{f} (z) \doteq \int\limits_{0}^{\infty} \frac{d w}{w} w^{-z} f(w).}
We will be concerned with finite diffusivity/viscosity $\nu$ in our equations, which amounts to $f(0) \doteq 1$ (neglecting normalization), and power law or faster decay at infinity with some so far unknown exponent $-\zeta$. The complex parameter $z$ in the above formula is therefore restricted to $-\mathcal R e (\zeta) < \mathcal R e (z) < 0$ \footnote{note that $\zeta$ may in fact be complex valued, in which case we would have $\zeta$ and $\zeta^*$ as the leading large scale exponents.}. The inverse transform is then
\eq{f(r) \doteq \int\limits_{c-\imath \infty}^{c+\imath \infty} \!\! \ud z \bar{f} (z) r^z}
where $-\mathcal R e (\zeta) < c <0$. Because of the constant boundary condition at zero, $\bar f (z)$ must have a pole at $z=0$. We will therefore take $c=0$ such that the contour will pass $z=0$ from the left, and denote this by $0-$ under the integration sign. We can now use the Mellin transform to \emph{define} a differential/integral operator of order $\sigma$ as
\eq{( \widehat D_\sigma f ) (r) \doteq \int_{0-}\!\! \ud z \Delta_\sigma (z) \bar f (z) r^{z-\sigma}.}
For example the derivative $\partial_r$ would correspond to $\sigma =1$ and $\Delta_1 (z) = z$. Consider now the equation
\eq{( \widehat D_\sigma f ) (r) + \lambda f (r) =0}
with the above mentioned boundary conditions. Expressing this with the help of the Mellin transforms yields
\eq{&&\int_{0-}\!\! \ud z \Delta_\sigma (z) \bar f (z) r^{z-\sigma} + \lambda \int_{0-}\!\! \ud z \bar f (z) r^{z} \nonumber \\ &&= \int_{0-}\!\! \ud z \Delta_\sigma (z) \bar f (z) r^{z-\sigma} + \lambda \int_{\sigma-}\!\! \ud z \bar f (z-\sigma) r^{z-\sigma} \nonumber \\ &&= \int_{0-}\!\! \ud z \left\{ \Delta_\sigma (z) \bar f (z) + \lambda \bar{f} (z-\sigma)\right\} r^{z-\sigma} + \sum_{i} r^{z_i} \mathcal R (\bar f (z_i)) \equiv 0,}
where in the second term on the second line we have simply changed the integration variable $z \to z-\sigma$ and shifted the contour from $\sigma \to 0$ on the third line (assuming sufficiently fast decay at imaginary infinities), and also denoted in the last sum the possible contribution of poles inside the strip $-\sigma < z < 0$. We can solve the problem by solving the recursion equation
\eq{ \Delta_\sigma (z) \bar f (z) + \lambda \bar{f} (z-\sigma) = 0 \label{recursion1}}
but \emph{only} if we can find a solution for which the strip of analyticity is pole free, i.e. that the sum of residues above vanishes, which also implies $\mathcal R e (\zeta) > \sigma$. It turns out that in the present context there are situations where such solutions cannot be found, at least not without improving the present procedure. We will however be able to find a sufficiently large class of solutions for which this problem does not appear.
\subsection{Isotropic $a$ -model equation for $\xi=2$}
For $\xi=2$ the problem becomes simple enough to be solved exactly
even for nonzero energies. From eq. (\ref{1overgamma}) we have now
\eq{\Lambda^a_2 (z) = (d-1) z^2 + \left( 4 a +d(d-1) \right) z + 2
a \left( d+ a (d^2-2) \right)}
and the equation (\ref{eigen}) can now be written as
\eq{\left( z-z_+ \right) \left( z-z_- \right) \bar g (z) = 2 \nu
\bar g (z+2) ,\label{xi2eq}}
where we have defined the roots
\eq{z_\pm = \frac{-4a -d (d-1) \pm \sqrt{D_E}}{2 (d-1)}}
with the discriminant
\eq{D_E = -4 (d-1) E +d^2 +d (d-2) \left( d^2 - 8 (d+1) a^2
\right).}
Employing the definition $\bar f (z) \doteq \mathcal A_z \bar g (z) \propto \frac{\Gamma\left( -z/2 \right)}{2^z \Gamma\left(\frac{z+d}{2} \right)} \bar g (z)$ and a translation $z \to z-2$, we obtain the equation
\eq{\frac{z(z+d-)}{(z-z_+-2)(z-z_- -2)} \bar f (z) + \frac{1}{2\nu} \bar f (z-2)=0,\label{xi2eq2}}
which should be compared to eq. (\ref{recursion1}). A general solution to eq. (\ref{xi2eq2}) can be written as
\eq{\bar f (z) = (2\nu)^{-z/2} \frac{\sigma_2 (z)}{\sin\left( \pi z/2 \right)}
\frac{\Gamma \left(
\frac{z-z_+}{2} \right)\Gamma \left( \frac{z-z_-}{2}
\right)}{\Gamma \left( \frac{z+2}{2} \right)\Gamma \left( \frac{z+d}{2} \right)}}
where $\sigma_2 (z+2) = \sigma_2 (z)$ is a so far arbitrary periodic function (with the subscript denoting the period). The solution (modulo the $\sigma_2$ -term) has poles at $z=2n$ and $z=z_\pm -2n$ for nonnegative integer $n$. The width of the strip of analyticity therefore has to be $-2 < \mathcal R e (z) <0$, from which we conclude that $\sigma_2(z)$ must be an entire, periodic function and that the poles $\mathcal R e (z_\pm) < -2$. At imaginary infinities the above solution behaves asymptotically as
\eq{\bar f (z) \sim  (2\nu)^{-z/2} \sigma_2 (z) e^{-\frac{\pi}{2} |z|} z^{a-1}.}
Since $\sigma_2 (z)$ is entire, we can expand it in Fourier series as $\sigma_2 (z) = \sum\limits_{k\geq 0} a_k \sin \left( k \pi z \right)$. Anything else than $k=0$ would however spoil the above asymptotic behavior, so we conclude that $\sigma_2 (z) = \text{const}$. Inverting the Mellin transform then yields
\eq{G(r) = C_1 \int_{0-} \!\! \ud z \frac{\Gamma \left(
\frac{z-z_+}{2} \right)\Gamma \left( \frac{z-z_-}{2}
\right)}{\Gamma \left( \frac{z+2}{2} \right)\Gamma \left( \frac{z+d}{2} \right)} \frac{\left( r^2 /2\nu\right)^{z/2}}{\sin\left( \pi z/2 \right)},}
which is what one would obtain e.g. in the MHD problem with $a=1$ by a direct coordinate space solution \cite{arponen}. The MHD problem for $\xi \neq 2$ can also be easily solved, although we will not reproduce that calculation here. The ground state energy is then the value of $E$ for which the discriminant $D_E$ vanishes (crossover between oscillating and power law decay), i.e.
\eq{E_0 = \frac{d \left( d^3 -10 d^2 + 9d +16 \right)}{4 (d-1)},}
which is negative for $3 \leq d \leq 8$ implying a dynamo effect for $\xi=2$. It is now tempting to use the above solution also for other values of $a$. However, for example for $a=-1$ and $d=3$ we have
\eq{z_\pm = -\frac{1}{2} \pm \frac{1}{2} \sqrt{-2 E -15}}
which means that for energies $E\geq -15/2$, the poles $z_\pm$ are always inside the strip of analyticity. We must therefore conclude that in cases such as this, the method is not sufficient to determine the existence of a steady state.

\newpage
\bibliography{bibsit}
\bibliographystyle{unsrt}

\end{document}